\documentclass{ita}
\usepackage{amssymb}
\usepackage{amsmath}
\usepackage{multirow}
\usepackage{wasysym}
\usepackage{textcomp}
\usepackage{enumerate}
\usepackage{comment}
\usepackage{color}
\newcommand{\dollar}[0]{\$}

\begin{document}

\title{Homing Vector Automata}\thanks{The first author is partially
supported by T\"{U}B\.{I}TAK (Scientific and Technological Research Council of Turkey).}
\author{\"{O}zlem Salehi}
\address{Bo\v{g}azi\c{c}i University, Department of Computer Engineering, Bebek 34342 Istanbul, Turkey \email{ozlem.salehi@boun.edu.tr \& say@boun.edu.tr}}
\author{A. C. Cem Say}
\sameaddress{1}
\author{Flavio D'Alessandro}
\address{Bo\u gazi\c ci University, 
Department of Mathematics, Bebek 34342, Istanbul, Turkey
\email{ dalessan@mat.uniroma1.it
	}}
\secondaddress{Universit\`a di Roma ``La Sapienza'', Dipartimento di Matematica, Piazzale Aldo Moro 2, 00185 Roma,  Italy }

\date{...}

\begin{abstract} 
We introduce homing vector automata,
which are finite automata augmented by a vector that is multiplied at
each step by a matrix determined by the current transition, and have to
return the vector to its original setting in order to accept the input. The
computational power and properties of deterministic, nondeterministic,
blind, non-blind, real-time and one-way
versions of these machines are examined and compared to
various related types of automata.  A generalized version of the
Stern-Brocot encoding method, suitable for representing strings on
arbitrary
alphabets, is also developed.
\end{abstract}

\maketitle
\section{Introduction}

The idea of augmenting the classical finite automaton model with an
external storage unit that can hold unlimited amounts of information,
yet can be accessed in a limited mode, is a celebrated topic of
automata theory, with pushdown automata \cite{Ch62} and counter
machines \cite{FMR67} as the most prominent examples.

Focusing on finite automata equipped with a register containing a
singleton, one can list automata with multiplication \cite{ISK76},
extended finite automata (EFA's) \cite{MS97} (also known as ``group automata"), and M-automata \cite{Ka09} among the
many such proposed models. In these machines, the register can
respectively store rational numbers, elements from a group, or a
monoid, and can be modified by multiplication. A computation is deemed
successful if the register, which is initialized to the identity
element, is equal to the identity element at the end.

Generalizing the idea of finite automata equipped with a register, we
have previously introduced \textit{vector automata} in \cite{SYS13}. A
vector automaton is a finite  automaton which is endowed with a
vector, and which can multiply this vector with an appropriate matrix
at each step. One of the entries of this vector can be tested for
equality to a rational number. The machine accepts an input string if
the computation ends in an accept state, and the test for equivalence
succeeds.

Many important models of probabilistic and quantum computation
\cite{Tu69,LR14} can be viewed in terms of vectors being multiplied by
matrices. Vector automata are useful for focusing on this matrix
multiplication view of programming, abstracting the remaining features
of such models away. In order to incorporate the aforementioned notion
of the computation being successful if the register/counter returns to
its initial value at the end of the computation to this setup,  we
propose the new  \textit{homing vector automaton} (HVA) model in this
paper. A homing vector automaton can multiply its vector with an
appropriate matrix at each step and can check the entire vector for
equivalence to the initial value of the vector. The acceptance
criterion is ending up in an accept state with the value of the vector
being equal to the initial vector.

We examine these machines under several different regimes, enabling us to determine the effect of definitional parameters such as whether the input is scanned in ``real time" or pausing the head on an input symbol for several steps is allowed, whether the machine can read its register during computation or is ``blind", with acceptance possible only if the register has returned to its initial value at the end, and whether nondeterminism confers any additional recognition power over deterministic programs.
We demonstrate a close relationship between the nondeterministic one-way blind variant
of the HVA model and the EFA's of \cite{MS97}, which we believe to be important for the following reasons.

The study of EFA’s until now essentially covered the cases of free
(non commutative) groups, and free abelian groups, together with their
algebraic extensions of finite index (virtually free groups), where some theorems of algebraic nature characterize the power of such models and the
properties of the languages recognized by these automata\cite{ISK76,DM00,Co05,Ka09}. 
There are no comparable general results for EFA's associated with groups other than the ones mentioned above.
In this theoretical setting, a model that seems natural to investigate is the linear one, that is, the one defined by  a group, or more generally, by  a semigroup of matrices over the field of rational numbers.  

Even in the cases of groups of matrices of low dimension (that are not of the types mentioned above), the study of HVA's and EFA's becomes quickly nontrivial, and there are remarkable classes of linear groups for which little is known about the EFA and HVA models that they define. The same consideration obviously holds for the more general case of machines defined by semigroups of matrices. 

Under this respect, the relationship among the two models exhibited here, and the fact that the new techniques (like the adaptation of the Stern-Brocot encoding method to HVA ``programming" in Section \ref{sec:SB}) in this paper can be ported to proofs about EFA's,  provide a new opening for EFA research. 

The rest of this paper is structured as follows: Section
\ref{sec:prel} contains definitions of basic terminology and the
machine models that will be compared to several restricted versions of
our model.
Section \ref{sec:hvadef} defines the homing vector automaton in its
most general (nondeterministic, one-way, non-blind) form, and
introduces the various limited versions that we will use to examine
the nature of the contribution of different aspects of the definition
to the power of the machine. In Section \ref{sec:ga}, we discuss the relationship between the nondeterministic one-way blind version
of the HVA model and the extended finite automata of \cite{MS97}, and use this
link to prove that these machines can recognize any Turing
recognizable language, even when the vector dimension is restricted to
four. We then focus on HVA's with real-time access to their input,
providing an exact characterization of the class of languages
recognized by these machines for the case where the alphabet is unary,
and showing that  the nondeterministic version is stronger than its
deterministic counterpart, recognizing some $ \mathsf{NP} $-complete
languages, in Section \ref{sec:rthva}.
A method we use for encoding strings on an alphabet of arbitrary size
in a blind homing vector automaton, based on Stern-Brocot trees
\cite{St58,Br61}, may be of independent interest. Section \ref{sec:SB}
contains a hierarchy result based on the dimension of the vector when
the matrix entries belong to a restricted set. Further results
regarding the model's relation with counter automata  and closure
properties are presented in Sections \ref{sec:ca} and
\ref{sec:closure}. Section \ref{sec:end} lists some open questions.

\section{Preliminaries}\label{sec:prel}

The following notation will be used throughout the paper: $Q$ is the set of
states, where $q_0 \in Q$ denotes the initial state, $Q_a \subset Q$ denotes the
set of accepting states, and $\Sigma$ is the input alphabet. $\Sigma^+$ denotes the set of all nonempty words
over $\Sigma$. An input string $w$ is placed between two endmarker symbols
on an infinite tape in the form 
$\cent w\dollar$. By $ w^r $, we represent the reverse of the string $ w $. $
w_i $ denotes the $ i $'th symbol of $ w $. The length of $ w $ is denoted by $ |w| $.

A machine can be real-time or one-way depending on the allowed tape head movements. If the tape head is allowed to stay put during some steps of its left-to-right traversal, then the machine is \textit{one-way}, and can make $\varepsilon $ (empty string) transitions without consuming any input symbol. A machine is \textit{real-time} if the tape head can only move to the right at each step. 

A machine M is said to \textit{recognize} a language \texttt{L}  if M  accepts all and only the members of \texttt{L}. For a machine model $A$, $\mathfrak{L}(A)$ denotes the class of languages  recognized by machines of type $A$.

Let $ K = (M,\circ, e )$ be a group under the operation denoted by $ \circ $ with the neutral
element denoted by $ e $. An \textit{extended finite automaton} \cite{DM00} over the group $ K = (M,\circ, e )$ (EFA($ K $)) is a
6-tuple
\[ \mathcal{E} = (Q, \Sigma,K,\delta, q_0,Q_a) \]
where  the transition function $\delta$ is defined as
\[\delta: Q \times (\Sigma \cup \{\varepsilon\}) \rightarrow \mathbb{P}(Q\times M).\] 
An extended finite automaton can be viewed as a nondeterministic finite automaton equipped with a register in which any element of $ M $ can be written. $ \delta(q,\sigma)=(q',m) $ means that when $\mathcal{E}$ reads the symbol (or empty string) $\sigma \in \Sigma \cup \{\varepsilon\}$
in state $q$, it will move to state
$q'$, and write $ x\circ m $ in the register, where $ x $ is the old content of the register. The initial value of the register is the neutral element $ e $ of the group $ K $. The string is accepted if after completely reading the string,  $\mathcal{E}$ enters an accept state, with the content of the register being equal to the neutral element of $ K $.

A \textit{real-time deterministic $k$-counter automaton} (rtD\textit{k}CA) \cite{FMR68} is a 5-tuple
\[ \mathcal{M}=(Q, \Sigma, \delta, q_0, Q_a). \]

The transition function $\delta$ of $\mathcal{M}$ is specified so that
$\delta(q,\sigma,\theta)=(q',c)$
means that $\mathcal{M}$
moves the head to the next symbol, switches to state $q'$, and updates its counters according to the list of increments represented by $c \in \{-1,0,1\}^k$,
if it reads symbol $\sigma \in \Sigma$, when  in state $q \in Q$, and
with $\theta  \in \{=,\neq\}^k $ describing whether the respective counter values equal zero
or not. At the beginning of the computation, the tape head is placed on the symbol $\cent$,
and the counters are set to 0. At the end of the computation, that is, after the right endmarker $\dollar$
has been scanned, the input is
accepted if $\mathcal{M}$ is in an accept state.

A \textit{real-time deterministic blind $k$-counter automaton} (rtD\textit{k}BCA) \cite{Gr78}
$\mathcal{M}$ is a D\textit{k}CA which can
check the value of its counters only at the end of the computation. Formally,
the transition function is now replaced by $\delta(q,\sigma)=(q',c).$ The input
is accepted at the end of the computation if $\mathcal{M}$ enters an accept
state, and all counter values are equal to 0.

\section{Homing vector automata}\label{sec:hvadef}
A \textit{one-way nondeterministic homing vector automaton} (1NHVA(\textit{k})) is a 6-tuple
\[\mathcal{V} =(Q,\Sigma,\delta,q_0,Q_a, {\bf v} ),\]
 where
\textbf{v} is a $k$-dimensional initial row vector, and the
transition function $\delta$ is defined as
\[\delta: Q \times
(\Sigma \cup \{\varepsilon \}) \times \Omega \rightarrow \mathbb{P}(Q\times S),\]
such that $\Omega=\{=,\neq\}$, where $ = $  indicates
equality to the initial vector $ {\bf v} $, and $ \neq $ otherwise, $\mathbb{P}(A)$ denotes the
power set of the set $A$, and $S$ is the set of $k \times k$ rational-valued matrices. The initial vector is freely chosen by the designer of the automaton.

Specifically, $(q',M) \in \delta(q,\sigma,\omega)$ means that when $\mathcal{V}$ consumes $\sigma \in \Sigma \cup \{\varepsilon\}$
in state $q$, with its current vector corresponding to $\omega \in \Omega$ ($\omega$ having the value = if and only if the current vector equals the initial vector),
it switches to state $q'$, multiplying its current vector with the matrix $M
\in S$ on the right. Thus the vector $  {\bf v}_i $ at step $ i $ is obtained by multiplying the vector $  {\bf v}_{i-1} $ at step $ i-1 $ by a specified matrix $ M $ so that $  {\bf v}_i= {\bf v}_{i-1}M $.
The string is accepted if $\mathcal{V}$ enters
an accept state, and the vector is equal to the initial vector $ {\bf v} $ as a result of arriving upon the right end-marker symbol
$\dollar$.

A \textit{one-way
nondeterministic blind homing vector automaton} (1NBHVA(\textit{k})) is a 1NHVA(\textit{k}) which is not allowed to check the vector until the end of the computation. The transition function $ \delta $ is defined as
$$ \delta: Q \times (\Sigma \cup \{\varepsilon\}) \rightarrow \mathbb{P}(Q\times S 
), 
$$
where $ (q',M) \in \delta(q,\sigma)$ means that when $\mathcal{V}$ consumes $\sigma \in \Sigma \cup \{\varepsilon\}$
in state $q$, it switches to state $q'$, multiplying its current vector with the matrix $M
\in S$ on the right. The acceptance condition is the same as for 1NHVA($ k $)'s.

A \textit{real-time deterministic homing vector automaton} (rtDHVA(\textit{k})) $ \mathcal{V} $ is a 1NHVA which is not allowed to make any nondeterministic moves and operates in real-time. The
transition function $\delta$ is defined as
\[\delta: Q \times \Sigma  \times \Omega \rightarrow Q\times S .\] 

A \textit{real-time deterministic blind homing vector automaton} (rtDBHVA(\textit{k})) is just a
rtDHVA(\textit{k}) which is not allowed to check the vector until the end of the computation. The transition function $\delta$ is now replaced by
$$ \delta: Q \times \Sigma \rightarrow Q\times S. $$

\section{Relationship with extended finite automata}\label{sec:ga}
In this section, we will exploit a relationship between 1NBHVA($ k $)'s and the extended finite automata of \cite{MS97} over free groups to demonstrate the power of homing vector automata. 

The two models seem to be linked in the case of extended finite automata over matrix groups, as the register is multiplied with a matrix at each step of the computation. Let us emphasize that the two models are different in the following sense. In a homing vector automaton, there is an initial vector \textbf{v}, and the accepted strings are those which label a computation path along which the product of the   sequence of matrices on the transitions is a matrix P, such that  ${\bf v}= {\bf v}$P. In the most general setting, the set of transition matrices belongs to the semigroup of rational matrices. In other words, in an accepting computation, the multiplied matrices belong to the stabilizer semigroup of the set of rational matrices with respect to $ \mathbf{v} $. In contrast, in an extended finite automaton over a matrix group, accepting computations are those in which the product of the transition matrices equals the identity matrix. In that sense, one-way nondeterministic blind homing vector automata can be seen as akin to what someone who wanted to define a version of EFA's associated with general matrix semigroups, rather than groups, would come up with. Some open questions regarding the link between the two models are listed in Section \ref{sec:end}.

We assume a familiarity of the reader with some basic notions from free group theory (see \cite{KM79,LS77} for classical references of this topic). Let us denote by $ \mathbf{F}_r $  the free noncommutative group over $r$ generators. Let us first recall some known results on such groups. A  well-known theorem by Nielsen and Schreier states that every subgroup of a free group is free (see \cite{LS77}, Proposition 2.11).  In particular, for every $r$ there is a set $X$ of $r$ elements so that the subgroup generated by $X$ is isomorphic to $ \mathbf{F}_r $.

We focus our attention on $ \mathbf{F}_2$. It is well known that $ \mathbf{F}_2$ admits a  representation by using matrices of
the group of all invertible matrices  of dimension $2$ over the ring of integers. In the sequel,  $id$ stands for the identity matrix.
Let $n$ be a positive integer and consider the group $K_n$ of matrices generated by  
\[ 
M_{a}=
\left [
 \begin{array}{rr}
 1&n\\
 0&1\\
 \end{array}
 \right ],~~~
 M_{b}=
\left [
 \begin{array}{rr}
 1&0\\
 n&1\\
 \end{array}
 \right ].
 \]

The following result holds (see \cite{KM79}, Theorem 14.2.1). 
\begin{fact}\label{thm: KM}
The group $K_n$ is isomorphic to  $ \mathbf{F}_2$. Moreover, if $  \mathbf{v} =\left [\begin{array}{rr}
	1&0
\end{array}\right ]$, for every 
matrix $M$ of $K_n$ which is not a power of $M_b$, $\mathbf{v} M \neq   \mathbf{v}.$
\end{fact}
As a straightforward consequence, there exists a subgroup $H$ of $K_n$ which is isomorphic to $ \mathbf{F}_2$ and such that:
\begin{equation}\label{eq:fla}
\forall \ M \in H \setminus \{id\}, \ 
 \mathbf{v} M \neq   \mathbf{v}. 
 \end{equation}
 Indeed, let $H$ be the subgroup of $K_n$ generated by $M_{a} M_{b} M_{a}^2$ and $M_{a}^2 M_{b} M_{a}$. 
 By the theorem of Nielsen and Schreier mentioned above, $H$ is freely generated by the latter two elements. In particular,  no element of $H$ equals a power of $M_b$. 
 This implies that (\ref{eq:fla}) holds for $H$. 
Denote
\begin{equation}\label{eq:flavar}
\varphi : \mathbf{F}_2 \rightarrow H,
 \end{equation}
 the  isomorphism  from $\mathbf{F}_2$ onto $H$.

Now we show that every extended finite automaton over a free group  can be simulated by a suitably defined homing vector automaton that is of dimension 2, nondeterministic, and one-way.  Precisely, we prove the following result. 
\begin{thrm}\label{thm: EFA-1NBHVA}
$ \mathfrak{L} \textup{(EFA$( \mathbf{F}_2)) $}  \subseteq \mathfrak{L}\textup{(1NBHVA(2))}.$
\end{thrm}
 \begin{proof}
 Let $\mathcal{E} = (Q, \Sigma, \mathbf{F}_2,\delta, q_0,Q_a)$ be an extended finite automaton on $ \mathbf{F}_2$. 
 Starting from $\mathcal{E}$, we construct a 1NBHVA(2) ${\cal V}= (Q, \Sigma, \mu, q_0,Q_a,  \mathbf{v})$ as follows. 
   Let $L$ be the finite set of elements of $\mathbf{F}_2$ defined as
 $$L= \{l\in \mathbf{F}_2: \exists \ p, q \in Q, \exists\ \sigma \in \Sigma \cup \{\varepsilon\} \ : \ (q, l)\in \delta (p, \sigma) \}.$$
 Set an enumeration on $L$ such that  $L = \{l_1, \ldots, l_r\},$ with $r\geq 1$ and let 
   $N= \{N_1, \ldots, N_r\}$, where, for every $i=1,\ldots, r$, $N_i=\varphi (l_i)$ is the image under the morphism (\ref{eq:flavar}) of $l_i$. The transition function of $\cal V$
 $$\mu : Q \times (\Sigma \cup \{\varepsilon\})  \rightarrow  \mathbb{P}(Q\times K_2)$$
 is defined as: for every $p, q\in Q$, and for every $\sigma \in \Sigma \cup \{\varepsilon\} $
 $$(q, l_i)\in \delta (p, \sigma) \ \Leftrightarrow (q, N_i) \in \mu (p, \sigma),$$
 where  $l_i\in L, N_i\in N$ and $N_i= \varphi(l_i)$.  Finally, we set $ \mathbf{v} =\left [\begin{array}{rr}
 1&0
 \end{array}\right ]$.
 
 Let $L_{\mathcal{E}}$ and $L_{\cal{V}}$ be the languages accepted by $\mathcal{E}$ and $\cal V$ respectively. Let us show that
 the two languages are equal. If $w= \varepsilon$, the claim is trivial. Suppose then 
 $w\in \Sigma ^+$.  If $w\in L_{\mathcal{E}}$, then there exists a computation of $\mathcal{E}$
 $$c = q_0 \stackrel{\sigma_1, g_1}{\longrightarrow}q_1  \stackrel{\sigma_2, g_2}{\longrightarrow} q_2 {\longrightarrow} \cdots  \stackrel{\sigma_k, g_k}{\longrightarrow}q_k,$$
from $q_0$ to a final state $q_k\in Q_a$ such that $w= \sigma_1 \cdots \sigma_k$ and  the element  $g\in \mathbf{F}_2$ associated with $c$  is $g= g_1\cdots g_k=e$.
By the definition of $\cal V$, there exists a computation of $\cal V$
 $$q_0 \stackrel{\sigma_1, M_1}{\longrightarrow}q_1  \stackrel{\sigma_2, M_2}{\longrightarrow} q_2 {\longrightarrow} \cdots  \stackrel{\sigma_k, M_k}{\longrightarrow}q_k,$$
such that, for every $i=1,\ldots, k$, $M_i = \varphi (g_i)$. Set $M= M_1\cdots M_k= \varphi (g_1) \cdots \varphi(g_k) = \varphi (g)$. Since $M=id$  we get ${\bf v}M = {\bf v}$, and $w\in L_{\cal{V}}$.

Suppose now that $w\in L_{\cal{V}}$. Then there exists a computation of $\mathcal{V}$
$$q_0 \stackrel{\sigma_1, {M}_1}{\longrightarrow}q_1  \stackrel{\sigma_2, {M}_2}{\longrightarrow} q_2 {\longrightarrow} \cdots  \stackrel{\sigma_k, {M}_k}{\longrightarrow}q_k,$$
from $q_0$ to a final state $q_k\in Q_a$, where $w= \sigma_1 \cdots \sigma_k$  and  the vector  associated with $c$  is $ {\bf v}M$, with $M= M_1\cdots M_k$. Since $w$ is accepted by $\cal V$, then $ {\bf v}M= \bf v$.
By (\ref{eq:fla}), then one has $M=id$. 
On the other hand, let the computation of ${\mathcal{E}}$
$$q_0 \stackrel{\sigma_1, {g}_1}{\longrightarrow}q_1  \stackrel{\sigma_2, {g}_2}{\longrightarrow} q_2 {\longrightarrow} \cdots  \stackrel{\sigma_k, {g}_k}{\longrightarrow}q_k,$$
where, for every $i=1, \ldots, k$, $\varphi (g_i)=M_i$. Then   the element $g=g_1\cdots g_k\in  \mathbf{F}_2$ is such that $M=\varphi (g)$. Hence $id=\varphi (g)$ implies $g=e$ and thus  $w\in L_{\mathcal{E}}$.
 \end{proof}

 \noindent
This allows us to draw the following conclusion about the class of languages recognized by 1NBHVA(2)'s.
 
 \begin{thrm}
 The family of context-free languages is included in $ \mathfrak{L} \textup{(1NBHVA(2)}). $ 
 \end{thrm}
 \begin{proof}
 Dassow and Mitrana \cite{DM00} provided (see \cite{Co05} and  \cite{Ka09} for alternative proofs that fix some details in the original proof) a characterization of context-free languages in terms of automata over a free group, namely, they stated that $ \mathfrak{L} \textup{(EFA$( \mathbf{F}_2)) $} $ is the family context-free languages.
The result then follows by Theorem \ref{thm: EFA-1NBHVA}.
 \end{proof}

Let $ \mathbf{F}_2 \times \mathbf{F}_2$ be the group given by the direct product of $ \mathbf{F}_2$ by $
\mathbf{F}_2$.
The following theorem characterizes the family of recursively enumerable languages.

\begin{thrm}\label{thm: EFA-RE} \textup{\cite{MS01}}
$ \mathfrak{L} \textup{(EFA$( \mathbf{F}_2 \times \mathbf{F}_2))$}$ is the family of recursively enumerable languages.
\end{thrm}

We can now demonstrate the huge power of 1NBHVA(4)'s. 

\begin{thrm}\label{thm: EFA-1NBHVA4}
  The family of recursively enumerable languages is included in $ \mathfrak{L}\textup{(1NBHVA(4))} $.
\end{thrm}
\begin{proof}
We will show how to simulate an EFA$( \mathbf{F}_2 \times \mathbf{F}_2)$ by a 1NBHVA(4). The result then follows from Theorem \ref{thm: EFA-RE}.

Let $\cal H$ be  the group of matrices of dimension $4$
\[ \left \{
\left [
 \begin{array}{clll}
\multicolumn{2}{c}{\multirow{2}{*}{{ } $M_1$}} & 0 & 0 \\
  & & 0 & 0 \\
 0&0 & \multicolumn{2}{c}{\multirow{2}{*}{{ } $M_2$}} \\
  0&0 & & \\
 \end{array}
 \right ], \ M_1, \ M_2 \in H \right \}.
  \] 
Since, by (\ref{eq:flavar}), $\varphi$ is an isomorphism  from $ \mathbf{F}_2$ onto   the group of matrices $H$, the mapping
 $\psi: \mathbf{F}_2 \times \mathbf{F}_2 \longrightarrow \cal H$ defined as:
 $$\forall \ (g_1, g_2)\in \mathbf{F}_2 \times \mathbf{F}_2, \  \psi(g_1, g_2) = (\varphi(g_1), \varphi (g_2)),$$ is an
  isomorphism from $ \mathbf{F}_2 \times \mathbf{F}_2$ onto $\cal H$.
  
  Let $\mathcal{E} = (Q, \Sigma, \mathbf{F}_2 \times \mathbf{F}_2,\delta, q_0,Q_a)$ be an extended finite automaton over $\mathbf{F}_2 \times \mathbf{F}_2$. 
 Starting from $\mathcal{E}$, we construct a 1NBHVA(4) ${\cal V}= (Q, \Sigma, \mu, q_0,Q_a,  \mathbf{v})$ as follows. 
   Let $L$ be the finite set of elements of $\mathbf{F}_2 \times \mathbf{F}_2$ defined as
 $$L= \{l\in \mathbf{F}_2 \times \mathbf{F}_2: \exists \ p, q \in Q, \exists\ \sigma \in \Sigma \cup \{\varepsilon\} \ : \ (q, l)\in \delta (p, \sigma) \}.$$
 Set an enumeration on $L$ such that  $L = \{l_1, \ldots, l_r\},$ with $r\geq 1$ and let 
   $N= \{n_1, \ldots, n_r\}$, where, for every $i=1,\ldots, r$, $n_i= \psi(l_i)$ is the image under the morphism $\psi$ of $l_i$. The transition function of $\cal V$
 $$\mu : Q \times (\Sigma \cup \{\varepsilon\}) \rightarrow  \mathbb{P}(Q\times {\cal H})$$
 is defined as: for every $p, q\in Q$, and for every $\sigma \in \Sigma \cup \{\varepsilon\}$
 $$(q, l_i)\in \delta (p, \sigma) \ \Leftrightarrow (q, n_i) \in \mu (p, \sigma),$$
 where  $l_i\in L, n_i\in M$ and $n_i= \psi(l_i)$.  Finally, we set $ 
 \mathbf{v} =\left [\begin{array}{rrrr}
  1&0&1&0
  \end{array}\right ]$.
 
 Let $L_{\mathcal{E}}$ and $L_{\cal{V}}$ be the languages accepted by $\mathcal{E}$ and $\cal V$ respectively. By using the very same argument of 
 the proof of Theorem \ref{thm: EFA-1NBHVA}, one verifies $L_{\mathcal{E}}=L_{\cal{V}}$.
\end{proof}

\section{Real-time homing vector automata}\label{sec:rthva}
   
   In the previous section, we have seen that allowing 
   one-way access to the input tape raises nondeterministic blind homing vector automata of small vector dimension to Turing equivalence. For this reason, we will be focusing on real-time input in the rest of the paper.

Another way in which one can examine the nature of the computational power of homing vector automata is by examining models in which the matrices used at each step for transforming the vectors are restricted in some way.
 Although the definition given in Section \ref{sec:hvadef} allows arbitrary rational matrices, we are going to constrain the matrix entries to belong to a particular set. In most automaton algorithms in this paper, the entries of the matrices belong to the set $ \{-1,0,1\} $, as this basic set will be seen to already capture many capabilities of homing vector automata. Let us note that multiplications with matrices whose entries belong to this set can be used to perform additions, subtractions, resets, and swaps between the vector entries. It is possible to recognize some of the languages in the following discussion with homing vector automata of lower dimension when a larger set of matrix entries is allowed. Some related open questions can be found in Section \ref{sec:end}.
  
  We start by comparing the deterministic blind and non-blind versions of our  model. 
  
\begin{thrm}\label{thm:blind}
$ \bigcup_k \mathfrak{L} 
\textup{(rtDBHVA(\textit{k}))} \subsetneq \bigcup_k \mathfrak{L} \textup{(rtDHVA(\textit{k}))}. $
\end{thrm}
\begin{proof}
It is obvious that any rtDBHVA($ k $) can be simulated by a rtDHVA($ k $). We are going to prove that the inclusion is proper by the witness language $ \mathtt{L}=\{a^nb^{a_1}a^{a_2}|n=a_1 \mbox{ or } n=a_1 + a_2\} $. Let us first construct a rtDHVA(2) $ \mathcal{V} $ recognizing $\mathtt{L} $. The idea is to simulate a counter with the help of the matrices. Starting with the initial vector 
$ 
\left [
\begin{array} {rr}
1&1\\
\end{array}
\right ]
 $, $\mathcal{V} $ multiplies the vector with the matrix $ M_+ $ for each  $ a $ it reads before the $b$'s,  incrementing the first entry of the vector with each such multiplication. After finishing reading the first segment of $ a $'s, $\mathcal{V} $ multiplies the vector with the matrix $ M_- $, decrementing the first entry of the vector for each $ b $.

$$M_{+}=
\left [
\begin{array} {rr}
1&0\\
1&1\\
\end{array}
\right ]~~~~~
M_{-}=
\left [
\begin{array}{rr}
1&0\\
-1&1\\
\end{array}
\right ]
$$ 

At each step, $ \mathcal{V} $ checks the current value of the vector for equality to $ 
\left [
\begin{array} {rr}
1&1\\
\end{array}
\right ]
$. If the equality is detected right after finishing reading the $ b $'s, it is the case that $ n=a_1 $, and $ \mathcal{V} $ multiplies the vector with the identity matrix at each step for the rest of the computation. If that is not the case, $ \mathcal{V} $ continues to multiply the vector with matrix $ M_- $ for each $ a $ after the $b$'s. The value of the vector will be equal to 
$  \left [
 \begin{array} {rr}
 1&1\\
 \end{array}
 \right ] $ at the end of the computation  if and only if $ n=a_1 $ or $ n =a_1+a_2$. 
 
Note that $ \mathtt{L} $ can be also recognized by a rtDHVA(1) by using the matrices $ M_{+}=2 $ and $ M_{-}=\frac{1}{2} $.
 
Now we are going to show that $ \mathtt{L} $ can not be recognized by any rtDBHVA($ k $). Suppose for a contradiction that $ \mathtt{L} $ is recognized by some rtDBHVA($ k $) $ \mathcal{V'} $. After reading a prefix of $ a $'s, the computation of $ \mathcal{V'} $ on a sufficiently long suffix of $b$'s will go through a sequence of states, followed by a state loop. Suppose that $ \mathcal{V'} $ is in the same state after reading two different strings $ a^nb^m $ and $ a^nb^n $, $ m<n $. Now consider the strings $u= a^nb^ma^{n-m} \in \mathtt{L} $ and $ w=a^nb^na^{n-m} \in \mathtt{L} $. After reading any one of these strings, $ \mathcal{V'} $ should be in the same accept state, and the  vector should be at its initial value. Assume that the strings in question are both extended with one more $ a $.  Since the same vector is being multiplied with the same matrix associated with the same state during the processing of that last $ a $, it is not possible for $ \mathcal{V'} $ to give different responses to $ 
a^nb^na^{n-m+1}$ and $ a^nb^ma^{n-m+1}$. Noting that $ a^nb^na^{n-m+1} \in \mathtt{L}$, whereas $ a^nb^ma^{n-m+1} \notin \mathtt{L}$, we conclude that $ \mathtt{L} $ can not be recognized by any rtDBHVA($ k $). 
\end{proof}

We can give the following characterization when the alphabet is unary.

\begin{thrm}\label{thm:unary}
For any $k$, all languages over $\Sigma = \{a\} $ accepted by a \textup{rtDHVA($ k $)} are regular.
\end{thrm}

\begin{proof}
Let $ \mathtt{L} $ be a unary language accepted by a rtDHVA($k$) $ \mathcal{V} $ and let $ {\bf v} $ be the initial vector of $ \mathcal{V}$. We are going to construct a DFA recognizing $ \mathtt{L} $ to prove that $ \mathtt{L} $ is regular. We assume that $ \mathtt{L} $ is infinite and make the following observation. Since $ \mathcal{V} $ has finitely many states, at least one of the accept states of $ \mathcal{V} $ will be accepting more than one string. Let $ w_1 $ and $ w_2 $ be the shortest strings accepted by an accept state $ q_a $ with $ |w_1|<|w_2| $. When accepting $ w_1 $ and $ w_2$, $ \mathcal{V} $ is in state $ q_a $ and the value of the vector is equal to $ {\bf v} $. After reading $ w_2 $, $ \mathcal{V} $ is in the same configuration as it was after reading $ w_1 $ and this configuration will be repeated inside a loop of $|w_2|-|w_1|= p $ steps. Therefore, we can conclude that all strings of the form $ a^{|w_1|+lp} $ for some positive integer $ l $ will be accepted by $ q_a $. 
 
Between consecutive times $ q_a $ accepts a string, some other strings may be accepted by some other accept states. Let $ u $ be a string accepted by $ q_b $ with $ |w_1| < |u| < |w_2| $. Then all strings of the form $ a^{|u|+lp} $ for some positive integer $ l $ will be accepted by $ q_b$ since every time $ \mathcal{V} $  enters the accepting configuration at state $ q_a $, $ \mathcal{V} $ will enter the accepting configuration at state $ q_b $ after $ |u|-|w_1| $ steps. The same reasoning applies to any other accepting configuration inside the loop. 
 
Now, let us construct a DFA $ \mathcal{D} $ accepting $ \mathtt{L} $. $ \mathcal{D} $ has $ |w_1|+1+(p-1) $ states. The first $ |w_1|+1 $ states correspond to the strings of length at most $ |w_1| $ and the state $ q_{|w|} $ is an accept state for all $ w \in \mathtt{L}$ that is of length at most $ |w_1| $. $ q_{|w_1|} $ and the next $ p-1 $ states $ q_{l_2},\dots,q_{l_p} $ stand for the configuration loop. States corresponding to accepting configurations inside the loop are labeled as accept states. 

The transitions of the DFA are as follows:
\begin{align*}
 \delta(q_i,a)&=q_{i+1} \mbox{ for }  i=0,\dots,|w_1|-1 \\
 \delta(q_{|w_1|},a)&=q_{l_2} \\
 \delta(q_{l_i},a)&=q_{l_{i+1}}  \mbox{ for }  i=2,\dots,p-1 \\
 \delta(q_{l_p},a)&=q_{|w_1|} \\
\end{align*}

Since $ \mathtt{L} $ can be recognized by a DFA, $  \mathtt{L} $ is regular. We conclude that any unary language accepted by a  \textup{rtDHVA($ k $)}  is regular.
\end{proof}

In the following theorem, we show that nondeterministic real-time homing vector automata are more powerful than their deterministic versions, both in the blind and nonblind cases.
\begin{thrm}\label{thm:upow}
\begin{enumerate}[i.]
\item $ \bigcup_k \mathfrak{L}\textup{(rtDBHVA(\textit{k}))} \subsetneq \bigcup_k \mathfrak{L} \textup{(rtNBHVA(\textit{k}))}$. 
\item $ \bigcup_k \mathfrak{L}\textup{(rtDHVA(\textit{k}))} \subsetneq \bigcup_k \mathfrak{L} \textup{(rtNHVA(\textit{k}))}$. 
\end{enumerate}
\end{thrm}
\begin{proof}
\textit{i.} It is obvious that a rtDBHVA($ k $) can be simulated by a rtNBHVA($ k $). We are going to show that the inclusion is proper by constructing a rtNBHVA(3) $ \mathcal{V} $ recognizing the unary nonregular language $ \mathtt{UPOW}'=\{a^{n+2^n}|n\geq 1\} $. Starting with the initial vector $  \left [
\begin{array} {rrr}
1& 1 & 1\\
\end{array}
\right ] $, $ \mathcal{V} $ multiplies the vector with matrix $ U_1 $ when reading each $ a $. The idea is to add the first and second entries together repeatedly to obtain powers of 2, so that after reading $ k $ symbols the value of the vector is equal to 
$  \left [
 \begin{array} {rrr}
  2^k&2^k&1\\
  \end{array}
\right ] $. $ \mathcal{V} $ nondeterministically guesses $ n $ and starts decrementing the first entry from that point on by multiplying the vector with the matrix $ U_2 $ which fixes the second entry to 1 immediately. At the end of the computation, the value of the vector is equal to 
$  \left [
   \begin{array} {rrr}
   1& 1 & 1\\
   \end{array}
   \right ] $ if and only if the input string is of the form $ a^{n+2^n} $ for some $ n $. 
   $$U_{1}=
   \left [
   \begin{array} {rrr}
   1&1&0\\
   1&1&0\\
   0&0&1\\
   \end{array}
   \right ]~~~~~
   U_{2}=
   \left [
   \begin{array}{rrr}
   1&0&0\\
   0&0&0\\
   -1&1&1\\
   \end{array}
   \right ]
   $$ 
From Theorem $ \ref{thm:unary}$, we know that every unary language recognized by a rtDHVA($ k $) is regular, concluding that $ \mathtt{UPOW}'\notin \bigcup_k \mathfrak{L} \textup{(rtDBHVA(\textit{k}))}$ .\\

\noindent \textit{ii.} It is obvious that a rtDHVA($ k $) can be simulated by a rtNHVA($ k $). The inclusion is proper as we have shown that $ \mathtt{UPOW}' $ can be recognized by a rtNHBVA(3), a feat that is impossible for rtDHVA($ k $)'s for any $k$.

\end{proof}

In the following theorem, we show that by allowing nondeterminism it is possible to recognize an $\mathsf{NP}$-complete language in real-time and with  matrices which are restricted to have integer entries. $\mathtt{SUBSETSUM}$ is the $\mathsf{NP}$-complete language which is the collection of all strings of the form $t \#  a_1\#...\# a_n\#$, such
that $t$ and the $a_i$'s are numbers in binary notation $(1 \leq i \leq n)$, and there
exists a set $I \subseteq \{1, . . . , n\}$ satisfying $\sum_{i \in
I}a_i=t$, where $n > 0$.  We define $\mathtt{SUBSETSUM}_r=\{ t^r \#  a_1^r\#...\# a_n^r\#\ | \exists I \subseteq \{1, . . . , n\} \mbox{ s.t. } \sum_{i \in
I}a_i=t\}$ in which the binary numbers appear in reverse order. It is obvious that $\mathtt{SUBSETSUM_r} \in \mathsf{NP}$, since $ \mathtt{SUBSETSUM}  \in \mathsf{NP} $. It is possible to reduce $\mathtt{SUBSETSUM}$ to $ \mathtt{SUBSETSUM_r}$ in polynomial time by reversing the binary numbers that appear in the input. Therefore, we can conclude that $ \mathtt{SUBSETSUM_r}$ is $\mathsf{NP}$-complete.   

\begin{thrm}
$ \mathtt{SUBSETSUM_r} \in \mathfrak{L}(\textup{rtNBHVA(5)}) $.
\end{thrm}
\begin{proof}
We construct a rtNBHVA(5) $\mathcal{V}$ recognizing 
$\mathtt{SUBSETSUM_r}$. The idea of this construction is to read the binary numbers in the string to entries of the vector, and to nondeterministically select the set of numbers that add up to $ t $.  We let the initial vector equal $ \left[
\begin{array}{rrrrr}
0&0&1&1&1
\end{array} \right ] $. We first encode $ t $ to the first entry of the vector as follows: While scanning the symbols of $t$, $\mathcal{V}$ multiplies the vector with the matrix $M_{T_0}$ (resp. $M_{T_1}$) for each scanned $0$
(resp. $1$). The  powers of 2 required for the encoding are obtained by adding the third and fourth entries, which always contain identical numbers, to each other, creating the effect of multiplication by 2. When $\mathcal{V}$ reads a $\#$, $\mathcal{V}$ multiplies the vector with the matrix $ M_{\#} $ which subtracts the second entry from the first entry and resets the second entry back to 0, and the third and  fourth entries back to 1.   
\[
 M_{T_0}=
\left[
\begin{array}{rrrrr}
1&0&0&0&0 \\
0&1&0&0&0\\
0&0&1&1&0\\
0&0&1&1&0\\
0&0&0&0&1
\end{array}
\right ]~~~~~
M_{T_1}=
\left[
\begin{array}{rrrrr}
1&0&0&0&0 \\
0&1&0&0&0\\
1&0&1&1&0\\
0&0&1&1&0\\
0&0&0&0&1
\end{array}
\right ]~~~~~
 M_{\#}=
 \left[
 \begin{array}{rrrrr}
 1&0&0&0&0 \\
 -1&0&0&0&0\\
 0&0&0&0&0\\
 0&0&0&0&0\\
 0&0&1&1&1
 \end{array}
 \right ] \]
 
In the rest of the computation, $\mathcal{V}$ nondeterministically decides which $a_i$'s to
subtract from the first entry.
Each selected $a_i$ is encoded using the same technique into the second entry of the vector. While scanning the symbols of $a_i$, $\mathcal{V}$ multiplies the vector with the matrix $M_{A_0}$ (resp. $M_{A_1}$) for each scanned $0$
(resp. $1$).

$$
M_{A_0}=
\left[
\begin{array}{rrrrr}
1&0&0&0&0 \\
0&1&0&0&0\\
0&0&1&1&0\\
0&0&1&1&0\\
0&0&0&0&1
\end{array}
\right ]~~~~~
M_{A_1}=
\left[
\begin{array}{rrrrr}
1&0&0&0&0 \\
0&1&0&0&0\\
0&1&1&1&0\\
0&0&1&1&0\\
0&0&0&0&1
\end{array}
\right ]
.$$

  $\mathcal{V}$ chooses another
$a_j$ if it wishes, and the same procedure is applied. At
the end of the input, $\mathcal{V}$ accepts if the vector is equal to 
$  \left[
 \begin{array}{rrrrr}
0&0&1&1&1 \\
 \end{array}
 \right ] $, which requires that the first entry of the vector is equal to 0. This is possible iff there exists a set of $ a_i $'s whose sum add up to $ t $.
\end{proof}

A language $ \mathtt{L} $ is in class $ \mathsf{TISP} $($ t(n),s(n) $) if there is a deterministic Turing machine that decides $ \mathtt{L} $ within $ t(n) $ time and $ s(n) $ space where $ n $ is the length of the input. Since the numbers in the vector can grow by at most a fixed number of bits in each multiplication, a Turing machine simulating a rtDHVA($ k $) requires only linear space \cite{SYS13}. Since the numbers in the vector can have length $O(n)$, whereas the matrix dimensions and entries are independent of the input length $n$, multiplication of a vector and a matrix requires $ O(n) $ time for each input symbol. We can conclude that $ \bigcup_k \mathfrak{L}$(rtDHVA($ k $))$ \subseteq  \mathsf{TISP}( n^2,n )$.

\section{Encoding strings with homing vector automata}\label{sec:SB}

\subsection{Stern-Brocot encoding}\label{sec:binary encoding}
The Stern-Brocot tree is an infinite complete binary tree whose nodes correspond one-to-one to positive rational numbers \cite{St58,Br61}. Crucially for our purposes, the Stern-Brocot tree provides a basis for representing strings as vectors of integers, as suggested for binary alphabets in \cite{GKP89}. 
The fractions in the Stern-Brocot tree can be stored as vectors of dimension 2, where the vector entries are the denominator and the numerator of the fraction. This representation allows us to perform the binary encoding easily in homing vector automata, as follows.

The empty string is represented by $ [\begin{array} {rr}
1&1
\end{array}] $. Now suppose that we want to encode a binary string $ w $ of length $ n $. For $ i=1 $ to $ n $, if $ w_i=0 $, we add the value of the first entry to the second one, and if $ w_i=1 $, we add the value of the second entry to the first one, multiplying the vector with the appropriate one of the following matrices $ M_0 $ and $ M_1 $:
$$ 
 M_{0}=
 \left [
 \begin{array}{rr}
 1&1\\
 0&1\\
 \end{array}
 \right ]~~~~~
 M_{1}=
 \left [
 \begin{array}{rr}
 1&0\\
 1&1\\
 \end{array}
 \right ]
 $$ 
A list of some binary strings and their encodings follows. A proof on the uniqueness of the encoding can be found in \cite{GKP89}. 

\begin{align*}
0 & \hspace{0.1in} 
[\begin{array} {lr}
1 & 2
\end{array}] 
& 00 & \hspace{0.1in}  [\begin{array} {lr}
1& 3
\end{array}] 
& 10 & \hspace{0.1in}  [\begin{array} {lr}
2& 3
\end{array}]  
& 000 & \hspace{0.1in}  [\begin{array} {lr}
1& 4
\end{array}]  & 010 & \hspace{0.1in}  [\begin{array} {lr}
3& 5
\end{array}]   \\
1 & \hspace{0.1in}  [\begin{array} {lr}
2& 1
\end{array}] & 01 & \hspace{0.1in}  [\begin{array} {lr}
3& 2
\end{array}]  & 11 & \hspace{0.1in}  [\begin{array} {lr}
3& 1
\end{array}]  & 001 & \hspace{0.1in}  [\begin{array} {lr}
4& 3
\end{array}]  & 011 & \hspace{0.1in}  [\begin{array} {lr}
5& 2
\end{array}]   
\end{align*}   
 
Given the vector representation $  {\bf v}_w $ of a string $ w $, it is also possible to decode the string with the following procedure: Let $ |w|=n $ and $  {\bf v}_w= [\begin{array} {lr}
a& b
\end{array}] $. Set $ w_n=0 $ if $ b>a $, and $ w_n=1 $ otherwise. Subtract the smaller entry from the larger one to obtain $  {\bf v}_w^{n-1} $ and repeat this routine until you obtain the vector  $ [\begin{array} {lr}
1& 1
\end{array}] $. When the given vector is not a valid representation of a string, then it is not possible to obtain  $ [\begin{array} {lr}
1& 1
\end{array}] $. The  matrices required for this procedure are $ N_0 $, which has the effect of subtracting the value of the first entry of the vector it is multiplied with from the second entry, and $ N_1 $, for the symmetric action. Note that $ N_{0} = M_{0}^{-1}$  and $ N_{1} = M_{1}^{-1} $.

$$ 
N_{0}=
 \left [
 \begin{array}{rr}
 1&-1\\
 0&1\\
 \end{array}
 \right ]~~~~~
N_{1}=
 \left [
 \begin{array}{rr}
 1&0\\
 -1&1\\
 \end{array}
 \right ]
 $$ 
%
%
  
\subsection{Generalized Stern-Brocot encoding}\label{sec:genstern}

We generalize the scheme mentioned above to strings on alphabets of arbitrary size and present a new method for encoding strings. Let $ \Sigma=\{a_1,a_2,\dots,a_k \} $, and $ w \in \Sigma^*$. With the \textit{generalized Stern-Brocot encoding} method described below, it is possible to uniquely encode $ w $ using a vector of size $ k $ and $ k \times k $ matrices whose entries belong to the set $ \{-1,0,1\} $. Let us note that one can use other methods to encode strings on arbitrary alphabet size using a vector of a smaller dimension but matrices whose entries belong to a larger set.   

 We start with the $ k $ dimensional vector $ [\begin{array} {rrrr}
 1&1&\dots & 1
 \end{array}] $, which represents the empty string. Suppose that  $|w|=n  $. To encode $w$, for $ i=1 $ to $ n $, if $ w_i=a_j $, the vector is multiplied with the  matrix $A_j  $, the $ k $ dimensional identity matrix whose $ j $'th column is replaced with a column of $ 1 $'s. Multiplication with $ A_j $ causes the  $ j $'th entry of the vector to be replaced by the sum of all the entries in the vector. 
 
 Among the different generalizations of the Stern-Brocot fractions, one that appears in \cite{Ga13} under the name of ``Stern's triatomic sequence'' is similar to the encoding we propose for the case $ k=3 $. The similarity lies in the construction of the sequence, but that sequence is not used for the purpose of encoding. As far as we know, no such generalization exists for the case $ k>3 $.
  
 In the following lemma, we prove the uniqueness of this generalized encoding.
 
 \begin{lmm}\label{lem:unique}
 No two distinct strings on $\Sigma$ \textup{ ($ |\Sigma|=k $)} can be represented by the same  vector of size $ k $ using the generalized Stern-Brocot encoding.
 \end{lmm} 
 \begin{proof}
 We will prove by induction on $n$ that if a $k$-dimensional vector $ {\bf v} $ is the generalized Stern-Brocot encoding of a string of length $n$, then $ {\bf v}$ is not the encoding of any other string of length at most $n$. 
  
 The empty string is represented by the $ k $-dimensional vector of 1's.
 The claim clearly holds for $n=0$, since no other strings of at most this length exist.
 Now assume that the claim holds for all natural numbers up to $ n-1 $. Let $ w $ be a string of length $ n $. The vector $  {\bf v}_w $ representing $ w $ is obtained by multiplying the vector $  {\bf v}_w^{n-1} $, representing the first $ n-1 $ symbols of $ w $, with $ A_j $ if $ w_n=a_j $. We will examine various possibilities regarding this final multiplication. Note that at a single step, it is possible to modify only a single entry of each vector. Now consider any string $ u \neq w $ with $ |u|=l $ and $ l \leq n $. If $ w $ and $ u $ have the same first $ n-1 $ symbols, then $  {\bf v}_w^{n-1}= {\bf v}_u^{l-1} $, the last symbols of the two strings are unequal, and it is not possible to obtain $  {\bf v}_w= {\bf v}_u $ since the same vector is multiplied by different matrices. In the remaining case, we know by the induction hypothesis that $  {\bf v}_w^{n-1}\neq  {\bf v}_u^{l-1} $. If these vectors disagree in more than two entries, there is no way that one can obtain the same vector by multiplying 
them once with some matrices of the form $A_j$. So we consider the case of the two vectors disagreeing in at most two entries.
 
 Suppose that $  {\bf v}_w^{n-1}$ and $ {\bf v}_u^{l-1} $ differ only in the  $ i $'th entry. If the final multiplications both work on the $i$'th entries, they will be adding the same number to them, resulting again in vectors differing in their $ i $'th entries. If one or more of the final multiplications deals with another entry, then the final vectors will surely disagree in that entry. It is not possible in any case to end up with equal vectors,
 
 Now suppose that $  {\bf v}_w^{n-1}$ and $ {\bf v}_u^{l-1} $ differ in two entries.
 If the final multiplications work on the same entry, then the final vectors will disagree in at least one entry.
 In the only remaining case, each one of the vectors is multiplied by a matrix updating a different one of the disagreeing entries. Let us represent the disagreeing entries of the vectors $  {\bf v}_w^{n-1} $ and $  {\bf v}_u^{n-1} $ by the pairs $(a, b)$ and $(c, d)$, respectively. Let $ x $ be the sum of the remaining $k-2$ entries in which the vectors agree. Without loss of generality, say that the entries become $(a, a+b+x)$ and $(c+d+x, d)$ after the final multiplication.
 But if the final vectors are equal, these pairs should also be equal, implying  $ c+b+2x=0 $, an impossibility. 
 
 We therefore conclude that it is not possible to have $  {\bf v}_w= {\bf v}_u $ for any string $ u $ of length at most $n$.
\end{proof}
 
Like in the binary case, given the vector representation  of a string, it is possible to reconstruct the string. The all-ones vector corresponds to the empty string. Any other vector $  {\bf v}_w$ encoding a string $w$ of length $n$ in this encoding has a unique maximum entry, say at position $j$. Then $ w_n$ is $a_j $, and we obtain $  {\bf v}_w^{n-1} $ by subtracting the sum of the other entries from the greatest entry. One repeats this procedure, reconstructing the string from right to left, until  one ends up with the all-ones vector. In terms of matrices, multiplications with the inverses of $ A_j $'s capture this process.  

\subsection{A hierarchy result}  
We will now use the generalized Stern-Brocot encoding to show a hierarchy result based on the dimension of the vector when an additional restriction is imposed on the matrices. 
 
\begin{thrm}\label{thm:hier}
Let $ S $ be the set of matrices whose entries belong to the set $  \{-m,-m+1,\dots,\allowbreak 0,\dots,m-1,m\} $ for some positive integer $m$, and let a \textup{rtDHVA($k$)} that is restricted to using members of $S$ in its matrices and initial vector be denoted a \textup{rtDHVA$_S$($k$)}. Then $ \mathfrak{L}\textup{(rtDHVA}_S(\textit{k})) \subsetneq \mathfrak{L} \textup{(rtDHVA}_S(\textit{l}))$ for $ l>(km)^k $. 
\end{thrm}

\begin{proof}
Using the generalized Stern-Brocot encoding, first we will show that it is possible to recognize $ \mathtt{MPAL_l}=\{w\#w^r|w\in\{a_1,a_2,\dots,a_l\}^*\} $ by a rtDHVA$_S$($ l $) $ \mathcal{V} $.
 
The input alphabet is $ \{a_1,a_2,\dots,a_l\} $,  and the corresponding matrices are \linebreak $ \{A_1,A_2,\dots,A_l\}, $ described in Section \ref{sec:genstern}. Starting with the $ l $ dimensional vector of 1's,  $ \mathcal{V} $ encodes the string by multiplying its vector with the matrix $ A_j $ whenever it reads an $ a_j $ until it encounters a $ \# $ . After reading the $ \# $, $ \mathcal{V} $ starts decoding by multiplying the vector with matrix $ A_j ^{-1}$ whenever it reads an $ a_j $.

If the string is of the form $ w\# w^r $, the vector will be multiplied with the inverse matrices in the correct order and the resulting value of the vector will be 
$ [\begin{array} {rrr}
 1&1 & \dots 1 
\end{array}]$. 

We also need to show that the input string is not accepted when it is not of the form $ w\#w^r $. Consider an input string $ x\#y^r $ and suppose that it is accepted by $ \mathcal{V} $. Let $ \bf v' $ denote the vector after reading $ x\# $ and let $ Y $ denote the product of the matrices the vector is multiplied while reading $ y^r $. Since the string is accepted, $ {\bf v'} Y=[\begin{array} {rrr}
1&1& \dots 1 
\end{array}] $ must be true. Since the matrices $ A_j^{-1} $ are invertible, $ Y $ is also invertible, which implies that $ \bf v' $ must be unique. Since $ y\#y^r \in \mathtt{MPAL}$, then $ \bf v' $ must be the vector obtained after reading $ y $ . From Lemma \ref{lem:unique}, we know that every string has a unique representation and we conclude that $ x $ and $ y $ are identical.

We are now going to show that $ \mathtt{MPAL_l}  \notin \mathfrak{L}(\textup{rtDHVA}_S(k))$ for $ l>(km)^k $. We first note that the value of any entry of a vector of size $ k $ can be at most $ m^{n+1}k^n $ after reading $ n $ symbols. This is possible by letting the initial vector  have $m$ in all entries, and multiplying the vector with the matrix with all entries equal to $ m $  at each step. Similarly, the smallest possible value of an entry is $ -m^{n+1}k^n  $, and so the number of possible different values for a single entry is $ 2m^{n+1}k^n+1 $.  If the machine has $ s $  states, $ s(2m^{n+1}k^n+1)^k $ is an upper bound for the number of different reachable configurations after reading $ n $ symbols. Since there are $ l^n $ strings of length $ n $ when the alphabet consists of $ l $ symbols, for large $ n $  and $ l >(km)^k $, the machine will end up in the same configuration after reading two different strings $ u $ and $ w $. This will cause the strings $ u\#w^r $ and $ w\#u^r $ which are not in  $ 
\mathtt{MPAL_l}$ to be accepted by the machine. Therefore, we conclude that $ \mathtt{MPAL_l} \notin \mathfrak{L}(\textup{rtDHVA}_S(k))$.

Since a vector automaton with a larger vector size can trivially simulate a vector automaton with a smaller vector size, the result follows.
\end{proof}

\section{Relationship with real-time counter automata}\label{sec:ca}

 A real-time deterministic homing vector automaton with a vector of dimension two can simulate a real-time deterministic one counter automaton (rtD1CA) which accepts with the condition that the counter is empty (See the proof of Theorem \ref{thm:blind}). The fact that the individual entries of the vector can not be checked prevents us from simulating a real-time deterministic multicounter automaton.

In the following theorem, we show that a rtDBHVA(2) can recognize a language which is not recognizable by any multicounter machine and we conclude that the language recognition powers of homing vector automata and multi-counter machines are incomparable. 
Note that the result also implies the incomparability of $ \bigcup_k \mathfrak{L} \textup{(rtDHVA($k$))} $ and $ \bigcup_k\mathfrak{L}\textup{(rtD\textit{k}CA)} $. This is not the case for the blind versions, as we prove in the second part of the theorem. 

\begin{thrm}\label{th:last}
\begin{enumerate}[i.]
\item $ \bigcup_k \mathfrak{L}\textup{(rtDBHVA($ k $))} $ and $ \bigcup_k \mathfrak{L} \textup{(rtD$ k $CA)} $ are incomparable.
\item $ \bigcup_k\mathfrak{L}\textup{(rtD$ k $BCA)} \subsetneq \bigcup_k \mathfrak{L} \textup{(rtDBHVA($ k $))}. $
\end{enumerate}
\end{thrm}
\begin{proof}
\label{ap:counter}
 \textit{i.} We know that $ \mathtt{MPAL_2}=\{w\#w^r|w\in\{0,1\}^*\} $ can be recognized by a rtDBHVA(2) by Theorem \ref{thm:hier}. In \cite{Pe11}, it is proven that no counter machine with $ k $ counters operating in time $ O(2^{n/k}) $ can recognize $ \mathtt{MPAL_2} $. Since we are working with real-time machines, the result follows.
 
 On the other hand, it is known that the nonregular unary language $ \mathtt{UGAUSS}=\{a^{n^2+n} | n \in \mathbb{N} \}$ can be recognized by a rtD2CA \cite{SYS13}. By Theorem \ref{thm:unary}, we know that rtDHVA($ k $)'s and inherently rtDBHVA($ k $)'s can recognize only regular languages in the unary case. Hence, we conclude that the two models are incomparable.
 
 \noindent \textit{ii.} Let us simulate a given rtD$ k $BCA $ \mathcal{M} $ by a rtDBHVA($ k+1 $). Let  $ [\begin{array} {rrrr}
 1&1&\dots & 1
 \end{array}] $
 be the initial vector of $ \mathcal{V} $. $ k+1 $'st entry of the vector will remain unchanged throughout the computation which will allow the counter updates. At each step of the computation, $ \mathcal{V} $ will multiply the vector with the appropriate matrix $ M\in S $ where $ S $ is the set of all $ (k+1)\times (k+1) $ matrices corresponding to possible counter updates. Since each counter can be decremented, incremented or left unchanged, $ |S|=3^k $. All matrices will have the property that $ M(i,i)=1  $ and $ M(k+1,k+1)=1 $. When the $ i $'th counter is incremented and decremented, then  $ M(k+1,i)=1 $ and $ M(k+1,i)=-1 $, respectively. At the end of the computation, the input will be accepted if the vector is equal to $ [\begin{array} {rrrr}
 1&1&\dots & 1
 \end{array}] $, which happens iff all counters have value 0.
   
 The inclusion is proper by the witness language $ \mathtt{MPAL_2} $.
\end{proof}

We have mentioned that deterministic blind homing vector automaton can recognize the language  $ \mathtt{MPAL_2} $ which is not recognizable by any counter machine. Consider the language  $ \mathtt{POW}=\{a^nb^{2^n}|n\geq 0\} $, whose Parikh image is not semilinear, which proves that the language is not context-free. Let us note that it is also possible to recognize $ \mathtt{POW} $ by a rtDBHVA(3) by using the same idea in the proof of Theorem \ref{thm:upow}.

\section{Closure properties}\label{sec:closure}

In this section, we examine the closure properties of the class of languages recognized by real-time homing vector automata. We start with a lemma which will be useful in  our proofs. The languages mentioned below are from \cite{ISK76}.

\begin{lmm}\label{lem: anb2n}
\begin{enumerate}[i.]
\item $ \mathtt{UNION}=\{a^nb^{n}|n \geq 0 \} \cup \{a^nb^{2n}|n \geq 0 \}  \notin \bigcup_k  \mathfrak{L} \textup{(rtDHVA(\textit{k}))} $.

\item $ \mathtt{L_{bab}}=\{b^n(a^nb^n)^k|n,k \geq 1\} \notin \bigcup_k  \mathfrak{L} \textup{(rtDHVA(\textit{k}))}  $ .

\item $ \mathtt{IJK}=\{a^ib^jc^k|i\neq j \mbox{ or } j > k\} \notin \bigcup_k  \mathfrak{L} \textup{(rtDHVA(\textit{k}))}  $ .

\item $ \mathtt{UNION_c}=\{a^nb^n|n\geq 0\} \cup \{a^nb^{2n}c|n\geq 0\}  \notin \bigcup_k  \mathfrak{L} \textup{(rtDHVA(\textit{k}))}  $ .
\end{enumerate}
\end{lmm}
\begin{proof}
We can show all these languages to be unrecognizable by rtDHVA's by applying the following common reasoning. Assume that the language $ \mathtt{L} $ in question is recognized by some rtDHVA($ k $)  $ \mathcal{V} $. Since there are finitely many states, one of the states of $ \mathcal{V} $ will end up accepting more than one member of the language. For each language, we will focus on two such members $u$ and $v$. Note that $ \mathcal{V} $ is in the same configuration (since it has also returned to its initial vector) after reading both $u$ and $v$. We then append another string $x$ to both strings, selected so that $ ux \in \mathtt{L} $ and $ vx \notin \mathtt{L}$. The responses of $ \mathcal{V} $ to the  $ ux $ and $vx$ has to be identical, since it will have returned to the same configuration after processing both strings. We conclude that $ \mathcal{V} $ can not distinguish between these two strings, and therefore that $ \mathtt{L} \notin \bigcup_k  \mathfrak{L} \textup{(rtDHVA(\textit{k}))} $. All that 
remains is to provide the strings $u$, $v$, and $x$ for the languages in the statement of the lemma. In the following, $i,j>1$ and $i\neq j$.

\begin{enumerate}[i.]
\item $u= a^ib^i $,  $v= a^jb^j $, and $ x=b^i $.

\item  $u= b^ia^ib^i $, $v= b^ja^jb^j $ and $x= a^ib^i $.

\item  $ u=a^ib^ic $, $ v=a^jb^jc $, and $x= c^{j-1}$ for $ i>j $.

\item $u= a^ib^i $, $v= a^jb^j $, and $x= b^ic $.
\end{enumerate}

\end{proof}

\begin{thrm}
\begin{enumerate}[i.]
\item $ \bigcup_k \mathfrak{L} \textup{(rtDHVA(\textit{k}))} $  is closed under the following operations:
\begin{enumerate}[a)]
\item intersection with a regular set
\end{enumerate}
\item $ \bigcup_k \mathfrak{L} \textup{(rtDHVA(\textit{k}))} $ is not closed under the following operations:
\begin{enumerate}[a)]
\item union
\item concatenation
\item intersection
\item star
\item homomorphism	
\item reversal
\item complementation
\end{enumerate}
\end{enumerate}
\end{thrm}
\begin{proof}\footnote{Let us note that it is possible to recognize the languages mentioned in the proofs with
rtDHVA($ k $)'s of smaller vector size when the vector entries are not restricted to be integers.}
\begin{enumerate}[i.]
\item 
\begin{enumerate}[a)]
\item Let $ \mathtt{L}_{\mathcal{V}} 
   $ be recognized by a rtDHVA($ k $)  $ \mathcal{V}= \newline (Q_1,\Sigma_1,\delta_1,q_1,Q_{a_1},v) $ and $
\mathtt{L}_{\mathcal{M}} $ be a regular language recognized by a finite state automaton $
\mathcal{M}=(Q_2,\Sigma_2,\delta_2,q_2,Q_{a_2}) $. Let us construct a rtDHVA($ k $) $
\mathcal{V'}=(Q,\Sigma,\delta,q_0,Q_a,v) $ recognizing $ \mathtt{L}=\mathtt{L}_{\mathcal{V}} \cap
\mathtt{L}_{\mathcal{M}} $. $ \mathcal{V'} $ keeps track of the vector and the current state of $ \mathcal{V}
$ as well as the current state of $ \mathcal{M}. $  Let $ Q' = Q_1 \times Q_2 $ be the state set of $
\mathcal{V'} $ and $ \Sigma=\Sigma_1 \cup \Sigma_2 $. For each $ (q_i,q_j) \in Q $, $ \sigma \in \Sigma $ and
$\omega \in \Omega$, $ \delta((q_i,q_j),\sigma, \omega)= ((q_i',q_j'),M)$ where $\delta_1(q_i,\sigma,
\omega)=(q_i',M)$ and $ \delta_2(q_j,\sigma)=q_j' $. $ q_0 $ is the pair $ (q_1,q_2) $ and $ Q_a $ is the set
of pairs of states where both of the states are accept states of $ \mathcal{V} $ or $ \mathcal{M} $. We 
obtain a rtDHVA($ k $) $ \mathcal{V}' $ recognizing $ \mathtt{L} $.
\end{enumerate}
\item \begin{enumerate}[a)]
\item Let $ \mathtt{L}_1=\{a^nb^n | n \geq 0\} $ and $ \mathtt{L}_2=\{a^nb^{2n}|n \geq 0\} $. $ \mathtt{L}_1 $ and $ \mathtt{L}_2 $ can be recognized by a rtDBHVA(2) which simulates a deterministic blind one-counter automaton whereas $  \mathtt{L}_1 \cup  \mathtt{L}_2 = \mathtt{UNION} $ can not be recognized by any rtDHVA($ k $) for any $ k $ by Lemma \ref{lem: anb2n}.

\item For the languages  $ \mathtt{L}_1=\{a^nb^n | n \geq 0\} $ and $ \mathtt{L}_2=\{a^nb^{2n}|n \geq 0\} $, $\mathtt{L}_1\mathtt{L}_2 \cap a^*b^*= \mathtt{UNION}  $,  which can not be recognized by any rtDHVA($ k $) for any $ k $ by Lemma \ref{lem: anb2n} and Part (i).a of this theorem.

\item Let $ \mathtt{L_1}=\{b^+(a^nb^n)^*| n \geq 1\} $ and $  \mathtt{L_2}=\{(b^na^n)^*b^+ | n \geq 1\} $. Both  $ \mathtt{L}_1 $ and $ \mathtt{L}_2 $ can be recognized by  rtDHVA(2)'s which simulate deterministic one-counter automata, whereas  $ \mathtt{L}_1 \cap \mathtt{L_2} = \mathtt{L_{bab}} = \{b^n(a^nb^n)^k|n,k \geq 1\}$ can not be recognized by any rtDHVA($ k $) for any $ k $ by Lemma \ref{lem: anb2n}.

\item Let $ \mathtt{L}=\{a^nb^n|n \geq 0 \} \cup \{ca^nb^{2n}|n \geq 0 \}$. A  rtDBHVA($ 2$) $ \mathcal{V} $ recognizing $\mathtt{L} $ branches into one of two computation paths depending on the first scanned symbol $ \sigma_1 $. If $ \sigma_1=a $, $ \mathcal{V} $ simulates a deterministic blind one-counter automaton recognizing $ \{a^{n-1}b^n | n \geq 0\} $ and if $ \sigma_1=c$, $\mathcal{V}$ simulates a deterministic blind one-counter automaton recognizing $\{a^nb^{2n}\}$. Now suppose $  \mathtt{L}^* \in \bigcup_k \mathfrak{L} \textup{(rtDHVA(\textit{k}))} $. Then $\mathtt{L}' = \mathtt{L}^* \cap \{ca^ib^j|i,j \geq 0\}= \{ca^nb^n|n\geq 0\} \cup \{ca^nb^{2n}| n \geq 0\} \in \bigcup_k \mathfrak{L} \textup{(rtDHVA(\textit{k}))} $. A rtDHVA($ k $) recognizing $ \mathtt{L}' $ can be easily modified to obtain a rtDHVA($ k $) recognizing the language $\mathtt{UNION} = \{a^nb^n|n\geq 0\} \cup \{a^nb^{2n}| n \geq 0\}   $, which is not in $ \mathfrak{L} \textup{(rtDHVA(\textit{k}))} $ by Lemma \ref{lem: anb2n}. 

\item Let $ \mathtt{L}=\{a^nb^n | n \geq 0\} \cup \{c a^{n-1}b^{2n}|n\geq 0\} $. A rtDBHVA($ k $) recognizing $ \mathtt{L} $ works similarly to the one in part d). Now consider the homomorphism $ h $ such that $ h(a)=a $, $ h(b)=b $ and $ h(c)=a $. $ h(\mathtt{L})=\{a^nb^n | n \geq 0\} \cup \{a^{n}b^{2n}\} = \mathtt{UNION}  $, which can not be recognized by any rtDHVA($ k $) for any $ k $ by Lemma \ref{lem: anb2n}.

\item Let $ \mathtt{L}=\{b^na^n | n \geq 0\} \cup \{cb^{2n}a^n|n\geq 0\} $. A rtDBHVA($ k $) recognizing $ \mathtt{L} $ works similarly to the one in part d). Now consider the reverse of $ \mathtt{L} $, $ \mathtt{UNION_c}= \{a^nb^n| n \geq 0\} \cup \{a^nb^{2n}c|n\geq 0\}$, which can not be recognized by any rtDHVA($ k $) for any $ k $ by Lemma \ref{lem: anb2n}.

\item Consider $ \mathtt{L}=\{a^mb^mc^n| 0\leq m \leq n\} $, which can be recognized by a rtDHVA(3). $ \bar{\mathtt{L}} \cap \{a^ib^jc^k |i,j,k \geq 0\} = \{a^ib^jc^k|i\neq j \mbox{ or } j > k\} =\mathtt{IJK}$ can not be recognized by any rtDHVA($ k $) by Lemma \ref{lem: anb2n}.
\end{enumerate}
\end{enumerate}
\end{proof}

The set of languages recognized by real-time nondeterministic homing vector automata is closed under union, star and concatention. The constructions are fairly simple and omitted. 

\begin{thrm}
\begin{enumerate}[i.]
\item $  \bigcup_k \mathfrak{L} \textup{(rtDBHVA(\textit{k}))} $ is closed under the following operations:
\begin{enumerate}[a)]
\item intersection
\end{enumerate}
\item 
$ \bigcup_k \mathfrak{L} \textup{(rtDBHVA(\textit{k}))} $ is not closed under the following operations:
\begin{enumerate}[a)]
\item union
\item concatenation
\item star
\item homomorphism
\item reversal	
\item complementation
\end{enumerate}
\end{enumerate}
\end{thrm}

\begin{proof}
\begin{enumerate}[i.]
\item 
\begin{enumerate}[a)]
\item Let $\mathtt{L}_{\mathcal{V}_1}$ and $ \mathtt{L}_{\mathcal{V}_2}$ be recognized by  rtDBHVA($ k_1$) \newline $ \mathcal{V}_1=(Q_1,\Sigma_1,\delta_1,q_1,Q_{a_1},v_1) $ and rtDBHVA($ k_2$) $ \mathcal{V}_2=(Q_2,\Sigma_2,\delta_2,q_2,Q_{a_2},v_2) $, respectively. Let us construct a rtDBHVA($ k $) $ \mathcal{V}=(Q,\Sigma,\delta,q_0,Q_a,v) $ recognizing $ \mathtt{L}=\mathtt{L}_{\mathcal{V}_1} \cap \mathtt{L}_{\mathcal{V}_2} $ where $ k=k_1 + k_2 $. Let $ Q = Q_1 \times Q_2 $ be the state set of $ \mathcal{V} $ and $ \Sigma=\Sigma_1 \cup \Sigma_2 $. For each $ (q_i,q_j) \in Q $ and $ \sigma \in \Sigma $, $ \delta((q_i,q_j),\sigma)= ((q_i',q_j'),M)$, where $\delta_1(q_i,\sigma)=(q_i',M_1)$, $ \delta_2(q_j,\sigma, \omega)=(q_j',M_2)$ and $ M $ is a $ k \times  k $ block diagonal matrix with $ M_1 $ and $ M_2 $ on its diagonal. $ q_0 $ is the pair $ (q_1,q_2) $, and $ Q_a $ is the set of pairs of states where both of the states are accept states of $ \mathcal{M}_1 $ or $ \mathcal{M}_2 $. The initial vector $ v 
$ of $ \mathcal{V} $ is of the form $ [v_1~v_2] $ and has dimension $ k $. $ \mathcal{V}$ keeps track of the current states and the current values of  both vectors by simultaneously multiplying its vector with the appropriate matrices. Since the computation is blind, the value of the vector is checked only at the end of the computation, and an input string is accepted if the vector is equal to its initial value.
\end{enumerate}
\item The proofs for the non-blind version also apply here. The proof for part f) follows from the fact that 
$ \bigcup_k \mathfrak{L} \textup{(rtDBHVA(\textit{k}))} $ is closed under intersection but not union.
\end{enumerate}
\end{proof}

The set of languages recognized by real-time nondeterministic blind homing vector automata is closed under union and intersection. The construction for union is straightforward, and the construction for intersection is identical to the deterministic case.

\section{Open questions}\label{sec:end}

What can we say about the relationship between real-time homing vector automata and one-way homing vector automata? We conjecture that one-way nondeterministic blind homing vector automata are more powerful than their real-time versions. Our candidate language is $ \mathtt{UPOW}=\{a^{2^n}|n\geq 0\} $, which can be recognized by a 1NBHVA(2). Note that when the machine in consideration is deterministic and blind, the real-time and one-way versions are equivalent in power. One can use the argument in Theorem 8 of \cite{SYS13} to prove this fact.

Can we show a separation result between the class of languages recognized based on the set of matrices used during the transitions of a homing vector automaton? Is it possible to recognize, for instance, the language $ \mathtt{POW_r}=\{a^{2^n}b^n|n\geq 0\} $ when the matrix entries are restricted to be integers? Note that it is possible to construct a rtDBHVA(2) recognizing $ \mathtt{POW_r} $ with the initial vector 
$  \left [
\begin{array}{rr}
  0&1
    \end{array} 
       \right ] $ 
    and the matrices $$ M_a=\left [
  \begin{array}{rr}
  1&0\\
  1&1\\
  \end{array}
  \right ] \mbox{ and } M_b=\left [
    \begin{array}{rr}
    \frac{1}{2}&0\\
    0&1\\
    \end{array}
    \right ] .$$
    
Can we show a hierarchy result between the classes of languages recognized by deterministic homing vector automata of dimensions $ k $ and $ k+1 $ for some $ k>1 $, maybe when the matrix entries are restricted to the set $ \{-1,0,1\} $? Consider the family of languages $ \mathtt{POW}(k)=\{ a^nb^{k^n}|n \geq 0\} $. We conjecture that it is not possible to recognize $ \mathtt{POW}(k)$ with a homing vector automaton of dimension less than $ k+1 $ with the restricted set of matrices.

Define a $ \textup{1NBHVA($ k $)}_G $ to be a $ \textup{1NBHVA($ k $)} $ where all transition matrices are elements of a matrix group $ G $. Can we always construct a $ \textup{1NBHVA($ k $)}_G $ recognizing the same language as a given EFA($G $)? (Note that we have proven that this is the case for $\textup{1NBHVA(2)}_{\mathbf{F}_2} $ and EFA($ \mathbf{F}_2) $.) 
Suppose that one can always find a suitable initial vector $ \textbf{v} $ such that for every $ M \in G $ except the identity matrix, $ \textbf{v}M\neq M $. Then one could construct the required  $ \textup{1NBHVA($ k $)}_G $ from the given EFA($G $) directly. For which groups $ G $ is it always possible to find such a vector?
	
What can we say about the reverse direction? For instance,  is every language recognized by some $ \textup{1NBHVA(2)}_{\mathbf{F}_2} $ necessarily in  $ \mathfrak{L} $(EFA($ \mathbf{F}_2 $))?

\section*{Acknowledgements}
We thank Ryan O'Donnell and Abuzer Yakary{\i}lmaz for their helpful answers to our questions, and the anonymous reviewers for their constructive comments.

\bibliographystyle{plain}
\bibliography{references}

\begin{thebibliography}{10}

\bibitem{Br61}
Achille Brocot.
\newblock {Calcul des rouages par approximation, nouvelle m\'{e}thode}.
\newblock {\em Revue Chronom\'{e}trique}, 3:186--194, 1861.

\bibitem{Ch62}
Noam Chomsky.
\newblock Context-free grammars and pushdown storage.
\newblock {\em M. I. T. Res. Lab. Electron. Quart. Prog. Report.}, 65:187--194,
  1962.

\bibitem{Co05}
Jon~M. Corson.
\newblock Extended finite automata and word problems.
\newblock {\em International Journal of Algebra and Computation},
  15(03):455--466, 2005.

\bibitem{DM00}
J\"{u}rgen Dassow and Victor Mitrana.
\newblock Finite automata over free groups.
\newblock {\em International Journal of Algebra and Computation},
  10(06):725--737, 2000.

\bibitem{FMR67}
Patrick~C. Fischer, Albert~R. Meyer, and Arnold~L. Rosenberg.
\newblock Real time counter machines.
\newblock In {\em Proceedings of the 8th Annual Symposium on Switching and
  Automata Theory (SWAT 1967)}, FOCS '67, pages 148--154, 1967.

\bibitem{FMR68}
Patrick~C. Fischer, Albert~R. Meyer, and Arnold~L. Rosenberg.
\newblock Counter machines and counter languages.
\newblock {\em Mathematical Systems Theory}, 2(3):265--283, 1968.

\bibitem{Ga13}
Thomas Garrity.
\newblock A multidimensional continued fraction generalization of stern’s
  diatomic sequence.
\newblock {\em Journal of Integer Sequences}, 16(2):3, 2013.

\bibitem{GKP89}
R.L. Graham, D.E. Knuth, and O.~Patashnik.
\newblock {\em Concrete mathematics: a foundation for computer science}.
\newblock Addison-Wesley, 1989.

\bibitem{Gr78}
S.~A. Greibach.
\newblock Remarks on blind and partially blind one-way multicounter machines.
\newblock {\em Theoretical Computer Science}, 7:311--324, 1978.

\bibitem{ISK76}
Oscar~H. Ibarra, Sartaj~K. Sahni, and Chul~E. Kim.
\newblock Finite automata with multiplication.
\newblock {\em Theoretical Computer Science}, 2(3):271 -- 294, 1976.

\bibitem{Ka09}
Mark Kambites.
\newblock Formal languages and groups as memory.
\newblock {\em Communications in Algebra}, 37(1):193--208, 2009.

\bibitem{KM79}
Mikhail~I. Kargapolov and Ju.~I. Merzljakov.
\newblock {\em Fundamentals of the Theory of Groups}.
\newblock Springer-Verlag, 1979.

\bibitem{LR14}
Richard~J. Lipton and Kenneth~W. Regan.
\newblock {\em Quantum Algorithms via Linear Algebra}.
\newblock MIT Press, 2014.

\bibitem{LS77}
Roger~C. Lyndon and Paul~E. Schupp.
\newblock {\em Combinatorial Group Theory}.
\newblock Springer-Verlag, 1977.

\bibitem{MS97}
Victor Mitrana and Ralf Stiebe.
\newblock The accepting power of finite automata over groups.
\newblock In {\em New Trends in Formal Languages}, pages 39--48.
  Springer-Verlag, 1997.

\bibitem{MS01}
Victor Mitrana and Ralf Stiebe.
\newblock Extended finite automata over groups.
\newblock {\em Discrete Appl. Math.}, 108(3):287--300, 2001.

\bibitem{Pe11}
Holger Petersen.
\newblock Simulations by time-bounded counter machines.
\newblock {\em International Journal of Foundations of Computer Science},
  22:395--409, 2011.

\bibitem{SYS13}
\"{O}zlem Salehi, Abuzer Yakary{\i}lmaz, and A.~C.~Cem Say.
\newblock Real-time vector automata.
\newblock In {\em Proceedings of the 19th International Conference on
  Fundamentals of Computation Theory}, FCT'13, pages 293--304. Springer-Verlag,
  2013.

\bibitem{St58}
Moritz~A. Stern.
\newblock \"{U}ber eine zahlentheoretische \mbox{F}unktion.
\newblock {\em Journal f\"{u}r die reine und angewandte Mathematik},
  55:193--220, 1858.

\bibitem{Tu69}
Paavo Turakainen.
\newblock Generalized automata and stochastic languages.
\newblock {\em Proceedings of the American Mathematical Society}, 21:303--309,
  1969.

\end{thebibliography}

\end{document}